\documentclass[onecolumn]{sdl}

\def\0{\mbox{\tiny $0$}}
\def\1{\mbox{\tiny $1$}}
\def\2{\mbox{\tiny $2$}}
\def\3{\mbox{\tiny $3$}}
\def\4{\mbox{\tiny $4$}}
\def\5{\mbox{\tiny $5$}}
\def\6{\mbox{\tiny $6$}}
\def\7{\mbox{\tiny $7$}}
\def\8{\mbox{\tiny $8$}}
\def\9{\mbox{\tiny $9$}}

\def\te{_\mathrm{TE}}
\def\tm{_\mathrm{TM}}

\usepackage{multirow}
\usepackage{color,colortbl}
\definecolor{bg}{rgb}{0.8,0.8,0.5}
\definecolor{bgA}{rgb}{0.98, 0.93, 0.36}

\logo{
\colorbox{DarkGoldenrod}{\color{white}$\mathbf{\Sigma\hspace*{0.06cm} \delta \hspace*{0.04cm} \Lambda}$}
}

\journal{\shadowtext{\textbf{\color{DarkRed} Laser Physics Letters}}\,\, \textbf{16}, 065001-5 (2019).}

\titlelines{3}
\title{Experimental evidence of laser power\\ oscillations induced by the relative\\ Fresnel (Goos-H\"anchen) phase}

\imgbgabstract{The amplification of the relative Fresnel (Goos-H\"anchen) phase by an appropriate number of total internal reflections and the choice of favorable incidence angles allow to observe full oscillations in the power of a DPSS laser transmitted
 through  sequential BK7 blocks. The experimental results confirm the theoretical predictions.  The optical apparatus used in this letter can be seen as a new type of two-phase ellipsometric system where the phase of the complex refractive index is replaced by the relative Fresnel (Goos-H\"anchen) phase.}

\author{
\names{S. A. Carvalho$^{1}$, S. De Leo$^{2,a}$, J. A. O. Huguenin$^{3}$, M. Martino$^{4}$, and L. da Silva$^{3}$}
\affiliation{$^{1}$Department of Exact Sciences, Fluminense Federal University, Volta Redonda, Brazil}
\affiliation{$^{2}$Department of Applied Mathematics, Campinas State University, Campinas, Brazil}
\affiliation{$^{3}$Department of Physics,  Fluminense Federal University, Volta Redonda, Brazil}
\affiliation{$^{4}$Department of Mathematics and Physics, Salento University, Lecce, Italy}
\email{$^{a}$deleo@ime.unicamp.br}
}

\begin{document}

\sdlmaketitle

Ellipsometry is a nonperturbing optical technique used for the characterization of surfaces, multilayers and
graded-index  films\,\cite{AB,HandB,Saleh}.  In the two-phase ellipsometric model,  a collimated beam of monochromatic light,  polarized in a known state, is  incident  upon  a single interface between  air and an isotropic media.  The   analysis of the reflected polarization state allows to calculate the ratio,  $\rho=R_{\tm}/R_{\te}$, between the transverse magnetic (TM) and electric (TE)  reflection coefficients,
\begin{equation}
\rho=\frac{\sin^{\2}\theta-\cos\theta\,\sqrt{n_*^{^{2}}-\sin^{\2}\theta}}{\sin^{\2}\theta+\cos\theta\,\sqrt{n_*^{^{2}}-\sin^{\2}\theta}}
=\tan\Psi\,\exp[i\,\Delta]\,\,,
\end{equation}
from which, once known the incidence angle $\theta$,
 the refractive index $n_*$ can be determined
\begin{equation}
n_*=|\sin\theta\,|\,\sqrt{1+\left(\frac{1-\rho}{1+\rho}\,\tan\theta\right)^{^{2}}}\,\,.
\end{equation}
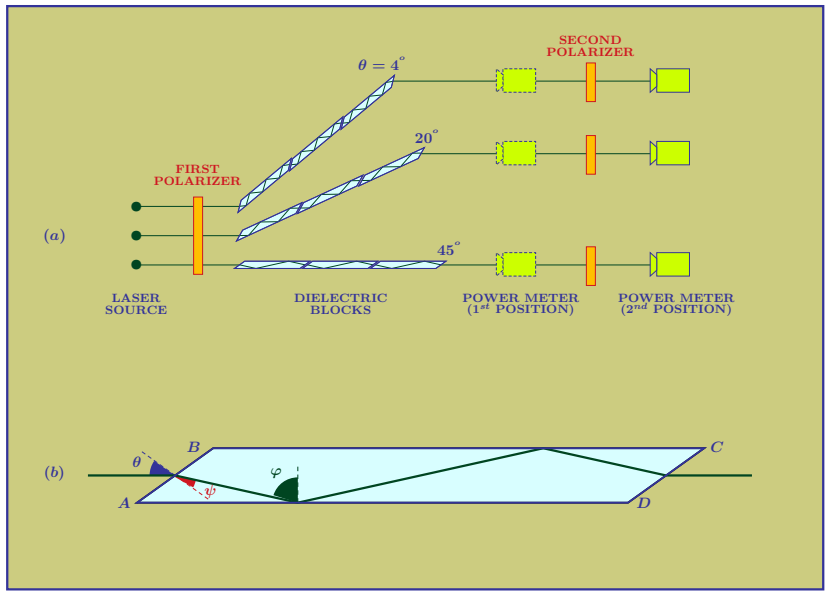
\WideFigureSideCaption{96-Fig1}{Experimental setup.  In (a), the electromagnetic radiation is emitted by a laser source and then mixed by polarizers (angle $\pi/4$) located before and after the dielectric structure. The experiment is done for three incidence angles: $4^{^{\mathrm o}}$, $20^{^{\mathrm o}}$, and $45^{^{\mathrm o}}$. The beam power is measured  before and after the second polarizer. In (b), the $x/z$ planar view of the BK7 block. Its geometry allows to give the refraction angle, $\psi$, and the incidence angle at the internal dielectric/air interface, $\varphi$, in terms of the incidence angle at the left air/dielectric interface, $\theta$.}
This approach assumes the absence of a transition layer or a surface film at the two-media interface.
In the two-phase ellipsometric model, the $\Delta$ phase is responsible for changing the linear polarization of the incident light into reflected elliptical polarization. A consequence of this phase is the possibility to induce  power oscillation
 by mixing the  TM and TE components of the reflected beam. This, for example, can be done by using a polarizer. Obviously, non absorbing materials, i.e. $\mathrm{Re}[n_*]=n$ and $\mathrm{Im}[n_*]=0$, imply  $\Delta=0$ and, consequently, the power oscillation  in the reflected beam can no longer be observed. As recently shown in ref.\,\cite{COL}, a dielectric block, with a real refractive index $n$, can be used as a new type of complex ellipsometric system, where,  as we shall see in detail later, the phase of the complex refractive index is replaced by
the phase which appears  in the Fresnel coefficient when the light is totally reflected at the internal dielectric/air interface.
This Fresnel phase is  responsible for the lateral displacement of light, named  Goos-H\"anchen  (GH) shift in honour of the physicists that in 1947\,\cite{GH47} provided, for TE waves, an experimental evidence of this phenomenon. In this letter, we shall refer to this phase as the Fresnel (Goos-H\"anchen) phase and use the abbreviation F(GH). After the Artmann theoretical analysis\,\cite{Art48} showing that lateral displacements can be calculated by using the derivative of the phase which appears in the Fresnel coefficient, the experiment was repeated, for TM waves, in 1949\,\cite{GH49} confirming the theoretical predictions done by Artmann one year before\,\cite{Art48}. The GH shift has been widely studied in literature\,\cite{rev1,rev2,rev3}. In the critical region, the behavior of light was recently theoretically studied and new phenomena investigated, see for example the axial dependence for incidence near the critical angle\,\cite{axial}, the breaking of symmetry and the consequent violation of the Snell law \cite{bs1,bs2} and, finally, the zitterbewegung  of light\,\cite{osc1,osc2}. Of particular interest is also the connection  between the delay time of quantum mechanics and the lateral shift in photonic tunneling\,\cite{dt1,dt2,dt3}. The frustrated total reflection\,\cite{exp1}, the axial dependence of  the composite Goos-H\"anchen shift\,\cite{exp2}, the transverse breaking of 
symmetry for transmission through dielectric slabs\,\cite{exp3}, and   the oscillatory behavior of light for critical incidence\,\cite{exp4} were also experimentally confirmed in recent investigations.

This great interest in lateral and angular shifts of light, stimulated a new analysis of the effect of the F(GH) phase on the propagation of a laser beam.  In particular, the complex GH ellipsometric system proposed in\,\cite{COL} allows, for an appropriate choice of the optical parameters,   to produce elliptic polarized light also in presence of a material with real refractive index and, thus,  to generate, by mixing the TM and TE components and amplifying the relative F(GH) phase, full oscillations in the power of the transmitted beam. In this letter, we experimentally test the
complex GH ellipsometric system proposed in\, \cite{COL} and  look for incidence angles and dielectric block geometries
for which a  full pattern of power oscillation and/or circular polarized light can be observed in the transmitted beam.
In presenting the experimental setup used to detect power oscillations, see Fig.\,1(a), we shall briefly recall the theoretical discussion which appears in ref.\,\cite{COL} adapting the formulas to the geometrical structure of our BK7 blocks, see Fig.\,1(b).

 A DPSS laser with $\lambda = 532.0\,\mathrm{nm}$,  power $1.5\,\mathrm{mW}$, and waist radius $\mbox{w}_{\0} = 79.0 \pm 1.5 \,\mu \mathrm{m}$ propagates along the $z$-axis passing through the first polarizer which selects linear polarized light with an angle $\pi/4$. A simple shorthand way to represent the incident polarization state is by the Jones vector\,\cite{Saleh}
 \begin{equation}
\label{eq:eqEinc}
   \boldsymbol{\mathcal{E}}_{{\mathrm{inc}}}(\boldsymbol{r}) = \left[ {\begin{array}{c}
   E_{{\mathrm x}}(\boldsymbol{r}) \\
   E_{{\mathrm y}}(\boldsymbol{r}) \\
  \end{array} } \right] = E(\boldsymbol{r}) \left[ {\begin{array}{c}
   1 \\
   1 \\
  \end{array} } \right]\,\,,
 \end{equation}
where

\begin{equation}
\label{eq:eqEin}
E(\boldsymbol{r}) = \displaystyle{\frac{E_{{\0}}}{\sqrt{1 + i\, z/z_{_{\mathrm R}}}}} \exp\left[- \frac{x^{^2} + y^{^2}}{\mbox{w}_{\0}^{^{2}}\,\left(1 + i\, z/z_{_{\mathrm R}}\right)}\right]
 \end{equation}
and $z_{_{\mathrm R}} = \pi \mbox{w}_{\0}^{^{2}} / \lambda$. The power of the incident beam is given by
\begin{equation}
   P_{\mathrm{inc}} = 2 \int \, dx \, dy \left|E(\boldsymbol{r})\right|^{^2}\,\,.
\end{equation}
Once passing through the first polarizer, the laser is transmitted through  dielectric structures composed by $N(=1,2,3)$  identical  BK7  ($n=1.5195$ at $\lambda=532\,\mathrm{nm}$) blocks. The dielectric structure is mounted on a high-precision rotation system which guarantees a  6 arcmin resolution when the dielectric system is rotated to fix the laser incidence angle $\theta$.
The F(GH) phase appears in the Fresnel coefficient when the light is totally reflected at the internal dielectric/air interface,  i.e. when $n\,sin\varphi\geq 1$ which implies $\varphi\geq 41.156^{\mathrm{o}}$ or equivalently $\theta\geq -\,5.847^{\mathrm{o}}$. For such incidence angles, the Fresnel transmission coefficients are  given by
\begin{equation}
T_{\sigma}= \left[\frac{4\,a_{\sigma}\,\cos\theta\,\cos\psi}{( a_{\sigma}\,\cos\theta\,+ \, \cos\psi)^{^2}}\right]^{^{N}}\, \exp[\,i\,\Phi_{\sigma}\,]\,\,,
\end{equation}
where $\sigma=\left\{\mathrm{TE}\,,\,\mathrm{TM}\right\}$, $\left\{a_{\te},a_{\tm}\right\} = \left\{1/n\,,\,n\right\}$ and
\begin{equation}
\Phi_{\sigma}= - \,2\,N_{\mathrm{ref}}\, \arctan \left[\,a_\sigma\,\frac{\sqrt{n^{2} \sin^{2}\varphi - 1}}{\cos \varphi}\,\right]
\end{equation}
is the F(GH) phase for TE and TM waves. We observe that in ref.\,\cite{COL} the F(GH) phase formula  is formally the same. Nevertheless, it is important to note that in ref.\,\cite{COL},  due to the different geometry of the dielectric block, $\varphi=\pi/2-\psi$ and this implies total internal reflection for incidence angles lesser than $\arcsin(\sqrt{n^{^{2}}-1})$, a  condition always respected for BK7 dielectric blocks. In our case, $\varphi=\pi/4+\psi$ and the F(GH) phase appears for incidence angles greater than $-5.847^{\mathrm{o}}$. The BK7 block dimensions are
\[
20.0\,{\mathrm{mm}}\,\,(\overline{AB})\,\, \times\,\,
91.5\,{\mathrm{mm}}\,\,(\overline{BC})\,\, \times\,\,
14.0\,{\mathrm{mm}}\,\,.
\]
For this  block we have 2 internal reflections for an incidence angle of $45^{\mathrm{o}}$, 4 reflections for $20^{\mathrm{o}}$, and 6 reflections for $4^{\mathrm{o}}$, see Fig.\,1(a).

The F(GH) phase is not the only phase appearing in the transmitted beam. The geometrical (or Snell) phase, $\Phi_{\mathrm{geo}}$, comes  from the continuity conditions at each air/dielectric and dielectric/air interfaces\,\cite{AJP}. Such a phase  is the same for TE and TM components and, due to the fact that we are interested in the relative phase between TE and TM components, we can omit its explicit expression. Thus, the laser transmitted through the dielectric structure composed by $N$ blocks, before passing through the second polarizer, is represented by
 \begin{eqnarray}
 \boldsymbol{\mathcal{E}}_{\mathrm{tra}}(\boldsymbol{r}) & \approx & \, E\left(\,x - x_{\mathrm{geo}}\,,\,y \,,\,z\,\right) \,\times\, \exp\left[i\,\Phi_{\mathrm{geo}}\right]\,\times \nonumber \\ \, && \left[ {\begin{array}{c}
 |T_{\tm}| \, \exp\left[i\,\Phi_{\tm}\right] \\
 |T_{\te}| \, \exp\left[i\,\Phi_{\te}\right]  \\
  \end{array} } \right]
\label{eq:eqEtra}
\end{eqnarray}
and its power by
\begin{eqnarray}
 P_{\mathrm{tra}} 	&=& \frac{|T_{\tm}|^{^{2}} + |T_{\te}|^{^{2}}}{2}\,\,P_{\mathrm{inc}}\,\,.
\end{eqnarray}
The second polarizer mixes the TE and TM components with an angle $\pi/4$ and, consequently, the transmitted field becomes

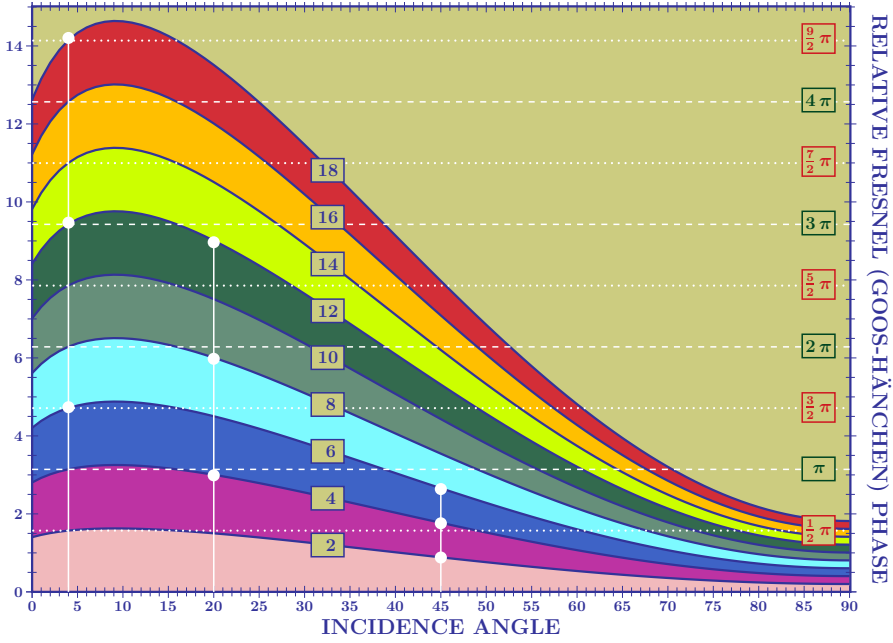
\WideFigureSideCaption{96-Fig2}{The relative F(GH)  phase is plotted for different number of internal reflections (2,4,...,18) as a function of the incidence angle. The vertical continuous lines are drawn in correspondence of the three incidence angles ($4^{\mathrm{o}}$,$20^{\mathrm{o}}$,$45^{\mathrm{o}}$) used to collect the experimental data. The white dots represent the relative F(GH) phase after 1, 2, and 3 BK7 blocks. The horizontal dashed lines intercept the F(GH) relative phase giving  inimal, $(1+2\,k)\,\pi$, and maximal, $2k\,\pi$, power and right, $(1/2+2\,k)\,\pi$, and left, $(3/2+2\,k)\,\pi$, polarized light.}

\begin{eqnarray}
 \boldsymbol{\mathcal{E}}_{\mathrm{pol}}(\boldsymbol{r}) & \approx & \, E\left(\,x - x_{\mathrm{geo}}\,,\,y \,,\,z\,\right) \,\times\, \exp\left[i\,\Phi_{\mathrm{geo}}\right]\, \times \nonumber \\ \, && \frac{|T_{\tm}| \, \exp\left[i\,\Phi_{\tm}\right] +  |T_{\te}| \, \exp\left[i\,\Phi_{\te}\right]}{2} \left[ {\begin{array}{c}
   1 \\
   1 \\
  \end{array} } \right]
 \end{eqnarray}
with  power given by
\begin{equation}
   P_{\mathrm{pol}} = \frac{|T_{\te}|^{^{2}} + |T_{\tm}|^{^{2}} + 2 \, |T_{\te} T_{\tm}|\, \cos \Phi_{_{\mathrm{F(GH)}}}}{4}\,
   P_{\mathrm{inc}}\,\,,
\label{Ppol}
\end{equation}
where
\begin{equation}
	\Delta_{_{\mathrm{F(GH)}}} = \Phi_{\te} - \Phi_{\tm} = 2 \,N_{\mathrm{ref}} \arctan \left[\, \frac{\sqrt{n^{2} \sin^{2}\varphi - 1}}{n \sin \varphi \tan \varphi} \right]\,\,,
	\label{phi_gh}
\end{equation}
is the relative phase between TE and TM components.
 In Fig.\,2, we draw the relative F(GH)  phase as a function of the incidence angle for different internal reflections,
\[N_{\mathrm{ref}}=2,\,4,\,...,\,18\,\,.\]
Fig.\,2 clearly shows that the relative F(GH) phase has to be amplified to obtain full oscillations in the power of the transmitted beam. This amplification can be achieved  by increasing the number of internal reflections, $N_{\mathrm{ref}}$. For example, for incidence at $\theta=20^{{\mathrm{o}}}$, the horizontal dashed line $\pi$ is very close to  the curve of the relative F(GH) phase corresponding  to 4 total internal reflections, this means practically minimal power at the exit of a single BK7 block. Obviously after 2 BK7 blocks we should find maximal power and, finally, after 3 BK7 blocks we have a power near its minimum value. The incidence angle $\theta=4^{\mathrm{o}}$ is of particular interest because it creates left polarized light after one BK7 block and right polarized light after 3 BK7 blocks. Indeed, the analytical solution  which guarantees, for 6 reflections,
$\Delta _{_{\mathrm{F(GH)}}}=3\pi/2$ is given for incidence incidence at $3.954^{\mathrm{o}}$. For such an incidence angle,
the  minimal power is thus reached after transmission through 2 BK7 blocks.

 The relative power, given by
\begin{equation}
\mathcal{P}_{\mathrm{rel}} = \frac{P_{\mathrm{pol}}}{P_{\mathrm{tra}}}=\frac{1 + \tau^{\,\2} + 2\, \tau \cos 	\Delta_{_{\mathrm{ F(GH)}}}}{2\, \left(1 + \tau^{\,\2}\right)}\,\,,
\label{eq:eqP}
\end{equation}
where
\[
\tau = \left|\frac{T_{\tm}}{T_{\te}}\right| =\left(\,\frac{\cos\theta + n\,\cos\psi}{n\,\cos\theta + \cos\psi}\,\right)^{^{2\,N}} \,\,,
\]
is what we aim to measure in our experiment. In Table\,\ref{tab:tab1}, we list the power of the transmitted beam before,
$P^{^{[\mathrm{exp}]}}_{\mathrm{tra}}$, and after, $P^{^{[\mathrm{exp}]}}_{\mathrm{pol}}$, the second polarizer for the incidence angles, $\theta = 4^{\circ}$, $20^{\circ}$, and $45^{\circ}$. The power measurement is repeated for all possible dielectric configurations, i.e. after one, two, and three aligned BK7 blocks.  The relative power,  $P^{^{[\mathrm{exp}]}}_{\mathrm{rel}}$, appears in the last column of Table\,\ref{tab:tab1}.

\begin{table}[htbp]
\centering
\begin{tabular}{ |>{\columncolor{bg}}c||>{\columncolor{bg}}c||rrr|rrr||>{\columncolor{bg}}r >{\columncolor{bg}}r>{\columncolor{bg}}r| }
\hline
\cellcolor{bg} $\theta$  & $N$ &
\multicolumn{3}{|c|}{$P^{^{[\mathrm{exp}]}}_{{\mathrm{tra}}} \,[\,\mu W\,]$} &
\multicolumn{3}{|c||}{$P^{^{[\mathrm{exp}]}}_{{\mathrm{pol}}}\, [\,\mu W\,]$} &
 \multicolumn{3}{|c|}{\cellcolor{bg} $10^{^3}{\mathcal{P}^{^{[\mathrm{exp}]}}_{{\mathrm{rel}}}}$}   \\ \hline \hline
            & $1$    &    $353.2$ & $\pm$ & $7.3$ &  $168.3$ & $\pm$ & $8.0$ &  $477$ & $\pm$ & $25$ \\
 $4^{^{o}}$   & $2$   &    $294.7$ & $\pm$ & $6.7$ &  $8.5$ & $\pm$ & $2.0$ &  $29$ & $\pm$ & $7$ \\
             & $3$   & $236.2$ & $\pm$ & $7.8$ &  $137.6$ & $\pm$ & $22.4$ &  $583$ & $\pm$ & $97$ \\
\hline \hline
             & $1$    &    $303.1$ & $\pm$ & $12.5$ &  $10.0$ & $\pm$ & $0.3$ &  $33$ & $\pm$ & $2$ \\
$20^{^{o}}$   & $2$   &    $260.5$ & $\pm$ & $17.6$ &  $246.4$ & $\pm$ & $19.8$ &  $946$ & $\pm$ & $99$ \\
             & $3$    & $220.8$ & $\pm$ & $8.5$ &  $10.2$ & $\pm$ & $1.0$ &  $46$ & $\pm$ & $5$ \\
						\hline \hline
             & $1$    &   $399.5$ & $\pm$ & $2.9$ &  $328.6$ & $\pm$ & $10.0$ &  $823$ & $\pm$ & $26$ \\
$45^{^{o}}$   & $2$   &  $354.0$ & $\pm$ & $6.5$ &  $134.5$ & $\pm$ & $4.7$ &  $380$ & $\pm$ & $15$ \\
             & $3$    & $304.2$ & $\pm$ & $2.8$ &  $12.1$ & $\pm$ & $1.3$ &  $40$ & $\pm$ & $4$ \\ \hline
\end{tabular}
\caption{Experimental data for the transmitted power before, $P^{^{[\mathrm{exp}]}}_{\mathrm{tra}}$, and after,
$P^{^{[\mathrm{exp}]}}_{\mathrm{pol}}$, the second polarizer for different incidence angles and dielectric blocks configurations. The relative power is given in the last column.}
  \label{tab:tab1}
 \end{table}

\noindent
The experimental data of the relative power can then be  compared with the theoretical prediction of Eq.\,(\ref{eq:eqP}). This is done in Fig.\,3 where theory and experiment show an excellent agreement. Finally, the F(GH) optical device, proposed in ref.\,\cite{COL} and tested in this letter, can be seen
as a new type of two-phase ellipsometric system
\begin{equation}
\rho_{_{\mathrm{F(GH)}}}=\frac{T_{\tm}}{T_{\te}}=\tan\Psi_{_{\mathrm{F(GH)}}}\,\exp[\,-\,i\,\Delta_{_{\mathrm{F(GH)}}}\,\,]
 \end{equation}
where  $\Psi_{_{\mathrm{F(GH)}}}=\arctan\,\tau$ and the phase coming from the complex nature of refractive index is replaced by the relative F(GH) phase.  In future investigations, it could be interesting to study the simultaneous effect of these two phases.\\

The authors thanks Fapesp, CNPq, Faperj, and INFN for the financial support.

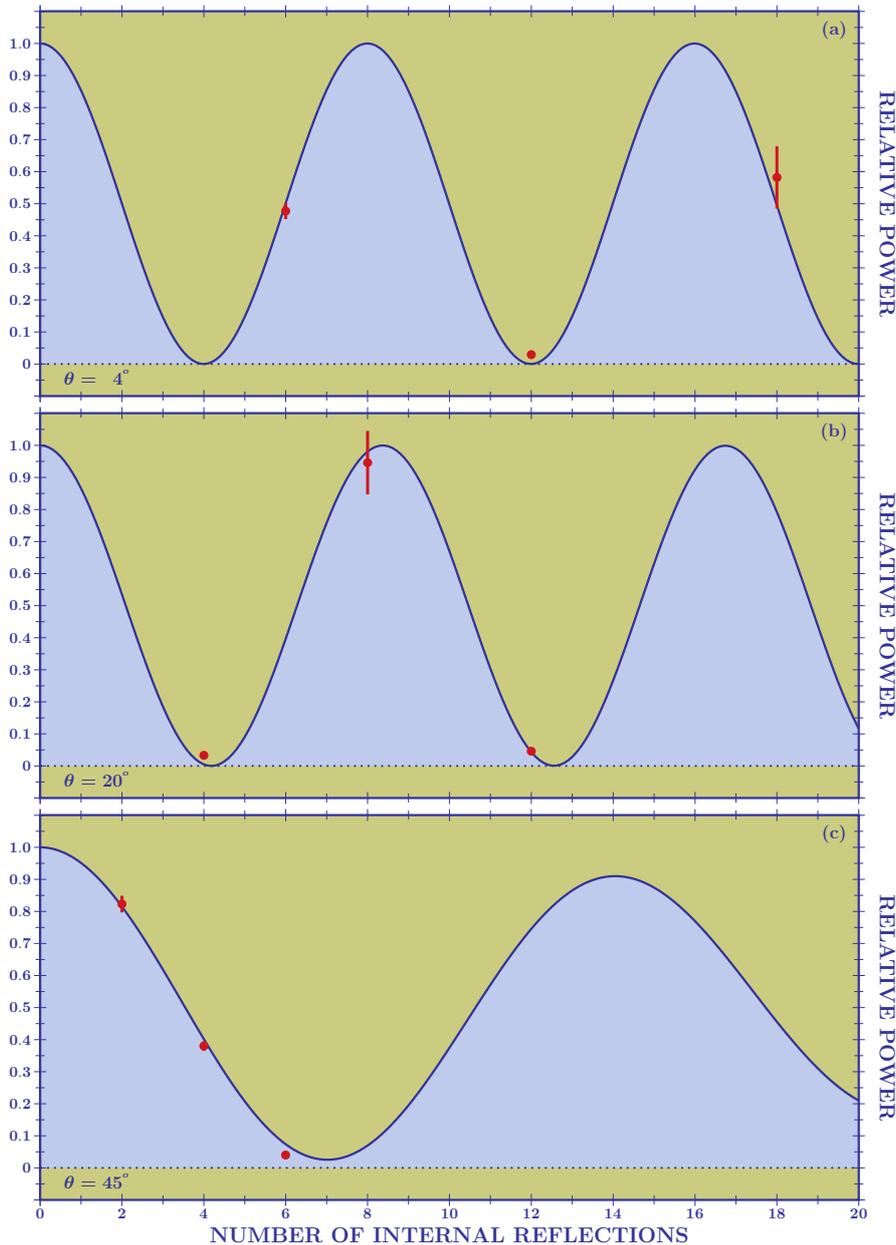
\WideFigureSideCaption{96-Fig3}{Experimental data (dots with error bars) and theoretical predictions (continuous lines) for the transmitted relative power for fixed incidences angles as a function of the number of internal reflections. For incidence at $\theta=4^{\mathrm{o}}$ the light transmitted through a single BK7 block suffers 6 internal reflections. For incidence at $\theta=20^{\mathrm{o}}$ and $45^{\mathrm{o}}$, we respectively have 4 and 2 internal reflections for each BK7 block.}

\end{document}